\documentstyle[12pt,a4,epsf]{article}
\newcommand{\rcite}[1]{{\cite{#1}}}
\newcommand{\rref}[1]{{(\ref{#1})}}
\newcommand{\tref}[1]{{\ref{#1}}}
\newcommand{\rlabel}[1]{{\label{#1}}}
\newcommand{\rbibitem}[1]{\bibitem{#1}}
\newcommand{\be}{\begin{equation}}
\newcommand{\ee}{\end{equation}}
\newcommand{\ba}{\begin{eqnarray}}
\newcommand{\ea}{\end{eqnarray}}
\newcommand{\dis}{\displaystyle}

\renewcommand{\mathrm}[1]{{\rm #1}}
\newcommand{\tr}{\mathrm{tr}}

\begin{document}
\begin{titlepage}
\begin{flushright}
{FTUV/97-7}\\
{IFIC/97-7}\\
{hep-ph/9702303}\\
\end{flushright}
\vspace{2cm}
\begin{center}
{\large\bf Kaon Polarizabilities in Chiral
Perturbation Theory\footnote{Work supported
in part by CICYT Grant No. AEN-96/1718 (Spain).}}\\
\vfill
{\bf Francisco Guerrero and Joaquim Prades}\\[0.5cm]
 Departament de
 F\'{\i}sica Te\`orica, Universitat de Val\`encia and\\
 IFIC, CSIC - Universitat de Val\`encia,
 C/ del  Dr. Moliner 50, \\ E-46100 Burjassot (Val\`encia),
Spain\\[0.5cm]
\end{center}
\vfill
\begin{abstract}
We study the kaon polarizabilities in the framework of
Chiral Perturbation Theory to order $p^4$. For the neutral
kaon we find that them vanish and they have the first
non-zero contribution  to order $p^6$. We also emphasize the
theoretical potential of an eventual measurement of
the kaon polarizabilities, in particular of the neutral kaon
ones.
\end{abstract}
\vspace*{1cm}
\vfill
February 1997
\end{titlepage}

 Electric and magnetic
polarizabilities are among the fundamental properties
of hadrons and provide valuable information on their internal
structure. They probe the rigidity of a composite
system against the formation
of electric (magnetic) dipole moments when an external
electric (magnetic) field is switched on. From an experimental
point of view,  it is possible to determine the polarizabilities
of a particle by  measuring its Compton scattering.
The influence of the polarizabilities on the $\gamma \gamma
\to P \overline P$ cross-section is small and one cannot
measure them very precisely, in general, from these analyses
\rcite{DH93}.

In this Letter we are interested in studying the
polarizabilities of kaons in the framework of Chiral Perturbation
Theory (CHPT) \rcite{WE,GL1,GL2}. For recent reviews on CHPT see
\rcite{rev}. For the pion polarizabilities,
the related process $\gamma \gamma
\to \pi \overline \pi$ has been studied at ${\cal O}(p^4)$ within
SU(3)$_L$ $\times$ SU(3)$_R$ CHPT in \rcite{BC} and in
\rcite{DHL} within SU(2)$_L$ $\times$ SU(2)$_R$,
and to order $p^6$ within SU(2)$_L$ $\times$ SU(2)$_R$
CHPT in \rcite{BGS} for the neutral pions and in \rcite{BU}
for the charged ones. The charged kaon polarizabilities
have been studied within CHPT in \rcite{DH89} to
${\cal O}(p^4)$. Though $\gamma \gamma \to
K \, \overline K$ processes are much beyond the applicability of CHPT
because of their center of mass energy,
one can expect CHPT to give a good description of
kaon polarizabilities within the usual 20\% because of
the SU(3) kaon mass breaking as we shall after see.
 The experimental situation
on kaon polarizabilities is very poor at the moment.
One could however expect an improvement in the future either
in the projected kaon factories like DA$\Phi$NE in Frascati
(see \rcite{DAFNE} for a thorough review of its physics
capabilities)
or in the high energy beam experiments at Fermilab and CERN
(see \rcite{BE,MO} and references therein).
We will see that in particular the neutral kaon polarizabilities
 can offer a nice way to test hadronic models
which predict related form factors in the kaon system.
This is particularly interesting since the same
models can be used to predict form factors for
radiative strangeness-changing processes like
the rare decay $K \to \pi \gamma \gamma$.

Expanding in photon momenta near threshold
the Compton amplitude for a pseudoscalar boson $P$
one can write down
\ba
\rlabel{first}
T\left(\gamma(q_1)P(p_1) \to \gamma(q_2) P(p_2)\right)
 \equiv \hspace*{4cm}
\nonumber \\ 2\left[ \vec {\epsilon_1} \vec {\epsilon_2^*}
 \left( e^2 - 4 \pi  \, m \, \overline \alpha \, \omega_1
\, \omega_2 \right)  - 4 \pi \, m \, \overline \beta
\, \left(\vec q_1 \times \vec {\epsilon_1} \right)
\left(\vec q_2 \times \vec {\epsilon_2^*} \right)
+ \cdots \right] \,
\ea
The phase convention we use can be obtained from this
amplitude  definition. Here $m$ is the pseudo-Goldstone boson
mass and $q \equiv (\omega, \vec q )$ and
$\epsilon  \equiv (0, \vec \epsilon )$ are photon momentum
and  polarization vector, respectively.
The Compton amplitude above can be decomposed in general
as follows
\ba
\rlabel{amp}
T\left(\gamma(q_1)P(p_1) \to \gamma(q_2) P(p_2)\right)
\equiv - e^2 A(t,\nu) \left[
(q_1 \cdot q_2) (\epsilon_1 \cdot \epsilon_2^*)
- (q_1 \cdot \epsilon_2^*) (q_2 \cdot \epsilon_1) \right]
\nonumber \\
- e^2 B(t,\nu) \left[ (q_1\cdot q_2) ( \Delta \cdot
\epsilon_1 ) (\Delta \cdot \epsilon_2^*) +
(\Delta \cdot q_1) ( \Delta \cdot q_2) (\epsilon_1
\cdot \epsilon_2^*) \right. \nonumber \\
\left. - (\Delta \cdot q_2) (q_1 \cdot
\epsilon_2^*) (\Delta \cdot \epsilon_1) -
(\Delta \cdot q_1) (q_2 \cdot \epsilon_1)
(\Delta \cdot \epsilon_2^*) \right]
\ea
for photons on-shell, where
\ba
s=(q_1+p_1)^2 \, ; \, t=(q_1-q_2)^2 \, ; \,
u=(q_1-p_2)^2 \, ; \, \nu \equiv s-u \, ; \,
\Delta \equiv p_1+p_2 \, .
\ea
For $p_1^2=p_2^2$ we have $2\Delta \cdot q_1
\, = \, 2 \Delta \cdot q_2 \, = \, s-u$.
The above amplitude is manifestly gauge invariant.

The polarizabilities $\overline \alpha$ and $\overline
\beta$ in \rref{first} can be obtained from the amplitudes defined
in \rref{amp} as follows
\ba
\rlabel{poldefs}
 \overline \alpha - \overline \beta
  &=& \frac{e^2}{4 \pi m} \, {\dis \lim_{t \to 0}} \,
\left(\overline A(t,\nu=t) + 8 m^2 \,
\overline B(t,\nu=t)\right) \nonumber \\
 \overline \alpha + \overline \beta
&=& \frac{e^2}{4 \pi m} \,
{\dis \lim_{t \to 0}} \,\left( m^2 \,
\overline B (t,\nu=t)\right) \,.
\ea
The barred amplitudes in \rref{poldefs} are the corresponding
amplitudes with the Born contributions using pseudoscalar
propagators to the corresponding order
 in CHPT  subtracted. The $\overline \alpha -
\overline \beta$ combination is pure S wave while
the $\overline \alpha + \overline \beta$ combination
is pure D wave. Here one observes that
polarizabilities are properties at $\nu=t \to 0$,
so the only possible parameters in
the chiral expansion are the ones which explicitly
break chiral symmetry, i.e.
masses of the pseudo-Goldstone
bosons. In known examples these corrections are
typically of the order of 20 \% to 30 \% in the
strange sector.
The point is that these type of
corrections find a natural framework within CHPT
so that CHPT for kaon polarizabilities is well
suited too.

At lowest order in CHPT, the only contribution to the
amplitudes $A(t,\nu)$ and $B(t,\nu)$ are the Born
type diagrams (see Figure \tref{f1})
\begin{figure}
\begin{center}
\leavevmode\epsfxsize=12cm\epsfbox{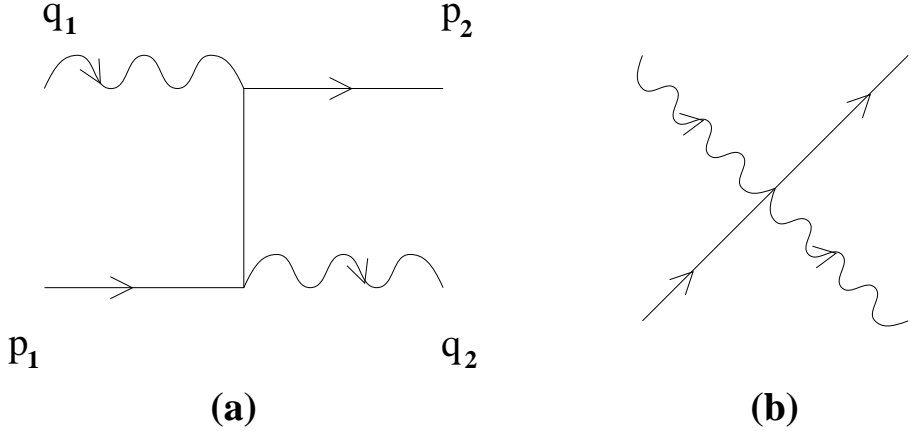}
\end{center}
\caption{\rlabel{f1}
U-channel diagrams contributing to $\gamma P \to
\gamma P$ at order $p^2$. S-channel
diagrams have to be included too. Straight lines are for
bosons and wavy for photons.}
\end{figure}
with the vertices from the  ${\cal O}(p^2)$ Lagrangian
\ba
\rlabel{p2}
{\cal L}^{(2)}= \frac{F_0^2}{4} \left\{
\tr \left(D_\mu U D^\mu U^\dagger\right) +
\tr\left(\chi U^\dagger + U \chi^\dagger\right)
\right\}
\ea
where $U \equiv {\rm exp} \left( \frac{\dis i
\sqrt 2 \Phi}{\dis F_0}  \right)$ is
an SU(3) matrix incorporating the octet of pseudoscalar mesons
\be \rlabel{Uoctet}
\Phi(x)=\frac{\vec{\lambda}}{\sqrt 2} \vec{\phi} =
\left( \begin{array}{ccc}
\frac{\dis \pi^0}{\dis \sqrt 2} + \frac{\dis \eta_8}
{\dis \sqrt 6} &  \pi^+ & K^+ \\
\pi^- & -\frac{\dis \pi^0}{\dis \sqrt 2}+\frac{\dis \eta_8}
{\dis \sqrt 6} & K^0 \\
K^- & \overline K^0 & -\frac{\dis 2 \eta_8}{\dis \sqrt 6} \end{array}
\right) \, .
\ee
In the absence of the U(1)$_A$ anomaly,
the SU(3) singlet $\eta_1$ becomes the ninth Goldstone boson which
is incorporated in the $\Phi(x)$ field as
\be
\Phi(x)= \frac{\vec \lambda}{\sqrt 2} \vec \phi +
\frac{\eta_1}{\sqrt 3}
\ee
Light-quark masses are collected
in the 3 $\times$ 3 flavor matrix ${\cal M}=
{\rm diag} (m_u,m_d,m_s)$ and $\chi \equiv 2 B_0 {\cal M}$.
The constant $B_0$ is related to the light-quark
vacuum expectation value
\be
\langle 0 | \overline q q | 0 \rangle =
-F_0^2 \, B_0 \left(1 + {\cal O}(m_q)\right) \, .
\ee
In this normalization, $F_0$ is the chiral limit value corresponding
 to the pion decay coupling $F_\pi \simeq 92.4$ MeV.
In the presence of electromagnetism
the covariant derivative $D_\mu$ is
\be
\rlabel{cova}
D_\mu U = \partial_\mu U -i |e| A_\mu \left[ Q ,
U \right] \,
\ee
Here, $A_\mu(x)$ is the photon field and the light-quark
electric charges in units of the electron charge $|e|$
are collected in the 3 $\times$ 3 flavor matrix $Q
={\rm diag} (2,-1,-1)/3$.

At next-to-leading order there are contributions from
the Born-type diagrams in Figure \tref{f1}
but with vertices from the order $p^4$ Lagrangian
(see \rcite{GL2} to find the explicit form)
and one-loop  diagrams in Figure \tref{f2}
and Figure \tref{f3} with vertices from the
${\cal O}(p^2)$ Lagrangian  in \rref{p2}.
\begin{figure}
\begin{center}
\leavevmode\epsfxsize=12cm\epsfbox{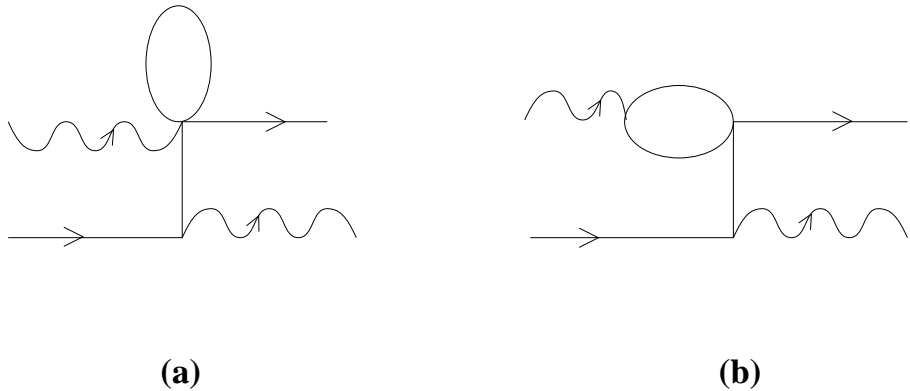}
\end{center}
\caption{\rlabel{f2} Pseudoscalar
electromagnetic form factor
diagrams contributing to $\gamma P \to
\gamma P$ at order $p^4$. S-channel and
symmetric diagrams have
to be  included too. Lines like in Figure
\protect{\tref{f1}}}
\end{figure}
\begin{figure}
\begin{center}
\leavevmode\epsfxsize=12cm\epsfbox{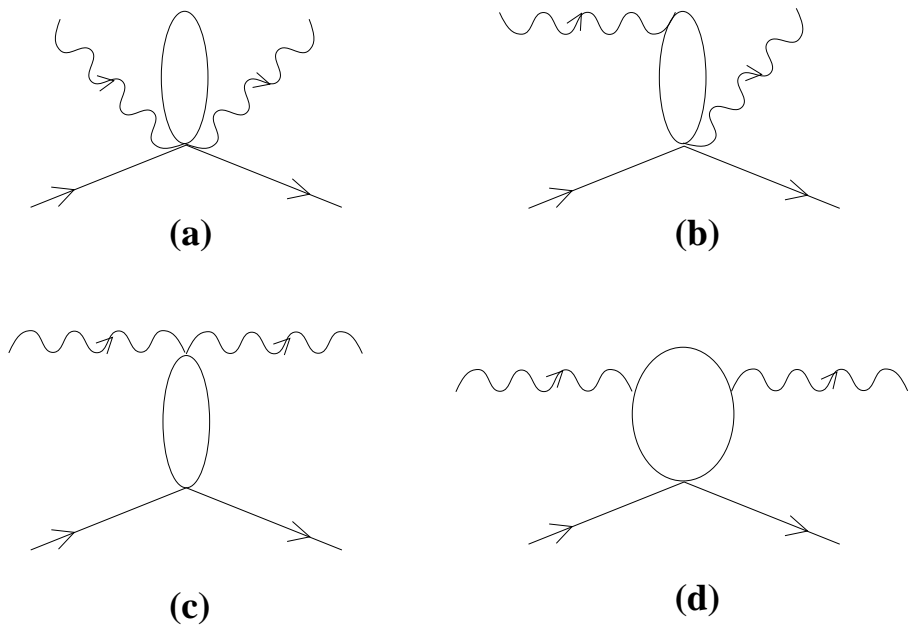}
\end{center}
\caption{\rlabel{f3} Diagrams contributing to $\gamma P \to
\gamma P$ at order $p^4$. Crossed diagrams
have to be included too. Lines like in Figure \protect{\tref{f1}}}
\end{figure}
The terms of the ${\cal O}(p^4)$ Lagrangian in \rcite{GL2}
are the needed counterterms (the so-called
$L_i$s couplings) to make the one-loop diagrams
with ${\cal O}(p^2)$ vertices UV finite.
For neutral pseudo-Goldstone  bosons, the first not
vanishing contribution to the Compton scattering
amplitude
come only from diagrams in Figure \tref{f3} \rcite{BC,DHL}.
 In fact, using the gauge invariance structure
of the Compton amplitude in \rref{amp}, one can reduce the
calculation of the ${\cal O}(p^4)$ contributions
for neutral pseudo-Goldstone bosons
to just diagram (d) in Figure \tref{f3}.
This is because the coefficients of the non
$(\epsilon_1 \cdot \epsilon_2^*)$
terms plus gauge invariance determine
uniquely the amplitudes $A(t,\nu)$ and $B(t,\nu)$ and
at this order only diagram (d) can generate non
$(\epsilon_1 \cdot \epsilon_2^*)$ terms.
Of course the result is finite. In addition
there are no counterterms to this order
so that the result we get
for the $K^0$ Compton scattering amplitudes up
to order $p^4$ in CHPT is
\ba
A(t,\nu) = -\frac{1}{16 \pi^2 F_0^2} \left[
2 - \frac{4 m_\pi^2}{t} \, \arctan^2
\left( \sqrt{\frac{t}{4 m_\pi^2-t}} \right)
\nonumber \right. && \\ \left.
- \frac{4 m_K^2}{t} \, \arctan^2
\left( \sqrt{\frac{t}{4 m_K^2-t}} \right) \right]
\, ; \nonumber &&\\ B(t,\nu)=0 \hspace*{7.5cm} \, .
\ea
The complete neutral kaon polarizabilities
to order $p^4$ is
\be
\rlabel{k0}
\overline \alpha_{K^0} = \overline \beta_{K^0}=0 \, .
\ee
Remember that the polarizabilities
for the $\pi^0$ at this same order obtained from the results
in \rcite{BC,DHL} are
\be
\overline \alpha_{\pi^0} = - \overline \beta_{\pi^0}=
- \frac{e^2}{4 \pi m_\pi} \, \frac{1}{96 \pi^2 F_\pi^2}
=-0.54 \, \times  \, 10^{-4}\, {\rm fm}^3 \,
\ee
where we have resummed the higher order corrections
that change $F_0$ into $F_\pi$.

For charged pseudo-Goldstone bosons,
there are order $p^4$  contributions
{}from the diagrams in Figures \tref{f1}, \tref{f2},
and \tref{f3}.
 In addition there are the diagrams which
give wave function and mass renormalization.
The complete result up to ${\cal O}(p^4)$
for the charged kaon  is the following
\ba
A(t,\nu) &=&
\frac{2}{t-\nu}+\frac{2}{t+\nu} +
\frac{8}{F_0^2}(L_9+L_{10}) \nonumber \\
&-& \frac{1}{16 \pi^2 F_0^2} \left[
\frac{3}{2} - \frac{2 m_\pi^2}{t} \, \arctan^2
\left( \sqrt{\frac{t}{4 m_\pi^2-t}} \right)
 \right. \nonumber \\
&-& \left. \frac{4 m_K^2}{t} \, \arctan^2
\left( \sqrt{\frac{t}{4 m_K^2-t}} \right) \right]  \, ;
\nonumber \\
B(t,\nu)&=& \frac{1}{t}\left[
\frac{1}{t-\nu} + \frac{1}{t+\nu} \right]\, .
\ea
This result agrees with the one found in \rcite{DH89}.
The corresponding complete charged kaon polarizabilities
to order $p^4$ are
\be
\rlabel{kplus}
\overline \alpha_{K^+} = - \overline \beta_{K^+}=
\frac{e^2}{4 \pi m_K} \, \frac{4}{F_K^2} (L_9+L_{10})
=(0.64 \pm 0.10) \,\times \,  10^{-4}\, {\rm fm}^3
\ee
where we have included the higher corrections
that change $F_0$ into $F_K$ and
used the recent result in \rcite{BT}.
\be
L_9+L_{10}=(1.6 \pm 0.2) \, \times \, 10^{-3}.
\ee

Let us now study the results we have obtained
in \rref{k0} and \rref{kplus}. Could we have gotten
them from some symmetry relation ? In the
$\gamma P \to \gamma P$ process,  the pseudoscalar boson $P$
has  to be in one of the three SU(2) subgroups of SU(3).
Then, unless the center of mass energy in
some channel is enough to
produce any of the rest of the octet multiplet
particles in \rref{Uoctet}, they can be integrated out
of the effective theory
and make the calculation within the corresponding
SU(2) CHPT. Of course in each one of these SU(2)
the coupling constants
depend on the ratio of the masses of the particles not integrated
over the ones of the integrated particles and
of $F_0^2$ over the masses of the integrated particles.
It has then little sense for numerical purposes
to integrate out the pion  in the case of calculations
involving kaons just because in nature we have
obviously not access to the effective couplings
of the U-spin SU(2) and the V-spin SU(2) CHPT.
It can nevertheless be useful for understanding some
results and/or getting new results.
The effective couplings corresponding to the SU(2)
isospin are known to ${\cal O} (p^4)$, they are the so-called
$l_i$s couplings defined in \rcite{GL1}.
For $m_P\neq0$ polarizabilities are in the situation
described above.

Therefore, formally  the calculation of pion polarizabilities
can be done in the isospin SU(2) CHPT, of charged
kaon within the V-spin SU(2) CHPT and of neutral kaons
within the U-spin SU(2) CHPT. Of course, the couplings in each
case will be different. Let us see what are the things we can
obtain from this observation. Since the quarks in the
u-d isospin system and in the u-s V-spin system
only differ for QCD in the presence of
electromagnetism  by the mass of the quarks the calculations
of charged kaon polarizabilities can be obtained from the
ones of charged pions by exchanging pion masses by kaon
masses. If we know
the relation between the order $p^4$ isospin SU(2) couplings
$l_i$s and the SU(3) couplings $L_i$s, we can have the result
for the charged kaons too just by translating the
$l_i$s couplings in the complete isospin SU(2) result
to SU(3) couplings and  changing pion by kaon masses.
The relation between $l_i$s and $L_i$s can be found in
\rcite{GL2,EGPR}. We find the result in \rref{kplus} without
any SU(3) calculation. Of course, some SU(3) calculation
was needed in order to make the relation between the SU(2)
and the SU(3) couplings, but these could be easier
(for instance two-point or three-point functions
calculations)  and they are  universal.

We cannot apply the same trick for the neutral
kaon polarizabilities because this SU(2) subgroup
 differs from the other
SU(2) subgroups also in the electric charge of the components
(s-d in this case). So the result in \rref{k0}
cannot be obtained from the neutral, charged
pion or charged kaon polarizabilities. This system has
the peculiarity in turn that we can only form electrically
neutral bosons. From this, we can easily obtain the result
in \rref{k0} as follows. Making the calculation
of the neutral kaon polarizabilities to ${\cal O}(p^4)$
in the U-spin
means that there are no loop contributions since there are
no photon--pseudo-Goldstone boson order $p^2$ vertices.
Noticing that there are no counterterms
in the U-spin CHPT calculation either
leads to the result in \rref{k0}. This is because
we get zero at ${\cal O}(p^4)$ in the complete (counterterm
plus loops) SU(2) calculation which
can only give zero when making the trick above explained
to go to SU(3). Therefore the vanishing result
for the neutral kaon polarizabilities at order $p^4$
is a result of chiral symmetry plus the fact that
the $K^0$ belongs to an SU(2) subgroup where there are
only neutral pseudo-Goldstone bosons.

One can see from the results in
\rcite{BC,DHL} that, at order $p^4$,
the kaon loops do not contribute to the pion polarizabilities.
This is just an accidental symmetry and it has not to happen
at higher orders. In general, there can be terms in the pion
polarizabilities that go to a constant  when $m_K \to \infty$
in the kaon loops of SU(3)$_L$ $\times$ SU(3)$_R$ CHPT.
These terms are included in the values of the $l_i$s
SU(2) couplings.

The order ${\cal O}(p^6)$ calculation of the charged pion
polarizabilities has been done in \rcite{BU}. We could use
the trick above again
to calculate the ${\cal O}(p^6)$ chiral log contributions
to the charged kaon from the charged pion calculation.
Here we also need the relation between the corresponding
order $p^6$ couplings for which one needs SU(3)
calculations. Notice that these
relations mix the SU(3) chiral loop contributions
with the SU(3) counterterms. The needed SU(3) calculations could be
easier however than $\gamma K^+ \to \gamma K^+$ itself
and as said before universal. For instance from two-point
and three-point function calculations to order $p^6$.
For the neutral kaon  polarizabilities,
the order $p^6$ relative
size cannot be guessed because its a complete new contribution
so both logs and counterterms to order $p^6$ have to be computed.
If calculated in the  U-spin CHPT,
the ${\cal O}(p^6)$ chiral logs for the neutral
kaons polarizabilities are again zero because
there neither photon--pseudo-Goldstone boson
order $p^4$ vertices, there are however non-zero
counterterm contributions of order $p^6$
this time. To get the corresponding
SU(3) result we need again the relation between the
corresponding order $p^6$ which however could be obtained
{}from easier SU(3) calculations as said before.

Let us finally analyze the possibilities that offer kaon
polarizabilities for studying the low-energy hadronic
interactions between pseudo-Goldstone bosons.

For pions and the charged kaon,
the combination of polarizabilities $\overline
\alpha - \overline \beta$ is not zero already at
order $p^4$ while the combination
$\overline \alpha+\overline \beta$ starts at order $p^6$.
Remember that many experimental
fits to pion polarizabilities are
made with the order $p^4$ constraint $\overline
\alpha + \overline \beta =0$; this has no sense
for the neutral kaon. For the neutral kaon we have obtained
that both combinations are first non-zero
at order $p^6$, so they are naturally
expected to be of the same order of magnitude.
This does not happen in any of the other systems.

The study of both the charged and the neutral kaon polarizabilities
is very interesting in relation with the information they can give
on the explicit breaking of chiral  symmetry through
kaon masses. As said before CHPT is the natural framework
to study such effects. In particular the neutral kaon
polarizabilities are proportional to $m_K$, so
the proportionality factor between the neutral kaon
polarizabilities  and $m_K$ is a direct measure of such
explicit chiral symmetry breaking effects.

The fact that the neutral kaon polarizabilities start
at order $p^6$  make them also very interesting for
checking  different hadronic models for counterterms.
In particular in checking the way they incorporate the explicit
chiral symmetry breaking effects.
In the case of the neutral kaon,
notice that SU(3) chiral symmetry together with
electromagnetism  forces the amplitudes $A(t,\nu)$
and $B(t,\nu)$ in \rref{amp} to go to zero when both
$m_K$ and $t$ go to zero. Any hadronic model has to
satisfy this constraint. For the neutral pions this is
only true in the large $N_c$ limit and $m_\pi$
and $t$ going to zero.

Calculations of the counterterms
appearing in charged and neutral kaon
polarizabilities can be found for instance in \rcite{BV}
using a Nambu--Jona-Lasinio
 model with no vector-like interactions, in
\rcite{BP} using an extended Nambu--Jona-Lasinio
 model with also spin-one
 interactions, in \rcite{IM} using the so-called
quark confinement model. The measurement of
kaon polarizabilities and in particular  the neutral kaon
ones can then test the predictability
for the order $p^6$ counterterms
of these models and others as vector meson dominance ones
 Here we want to emphasize that these
checks are only meaningful if a full CHPT calculation
(i.e. chiral logs and counterterms) at order $p^6$
is made since
we have no way of estimating their relative weight.
The analysis of neutral kaon polarizabilities
can also provide very useful information that
could be used in predicting some rare kaon decays form
factors for instance. We find then the neutral kaon
polarizability very interesting theoretically and deserving
further experimental effort at the planned kaon
facilities and experiments.

We want to thank Toni Pich for useful comments on the manuscript.
One of us (F.G.) acknowledges support from a FPI scholarship
of the Spanish Ministerio de  Educaci\'on y Cultura.

\end{document}